\documentclass[twocolumn]{aastex6}

\usepackage{mathtools}
\usepackage{enumitem}

\shorttitle{Extinction Correction significantly influences the estimate of \Lya\ escape fraction}
\shortauthors{An et al.}

\begin{document}
\definecolor{purple}{RGB}{160,32,240}
\newcommand{\peter}[1]{}
\newcommand{\fa}{F_\mathrm{H\alpha}}
\newcommand{\hinv}{h^{-1}}
\newcommand{\mpc}{\rm{Mpc}}
\newcommand{\hmpc}{$\hinv\mpc$}

\newcommand{\sfr}{\mathrm{SFR}}
\newcommand{\Msun}{M_{\odot}}
\newcommand{\Ha}{H$\alpha$}
\newcommand{\Hb}{H$\beta$}
\newcommand{\Lya}{Ly$\alpha$}
\newcommand{\HI}{{H}~{\scriptsize{I}}}
\newcommand{\OIII}{[{O}~{\scriptsize {III}}]}
\newcommand{\HII}{{H}~{\scriptsize {II}}}
\newcommand{\CIV}{{C}~{\scriptsize{IV}}}
\newcommand{\OII}{[{O}~{\scriptsize {II}}]}
\newcommand{\SIII}{[{S}~{\scriptsize {III}}]}
\newcommand{\SII}{[{S}~{\scriptsize {II}}]}
\newcommand{\C}{$\kappa({\rm L})$}
\newcommand{\mm}{$\mu$m}

\newcommand{\phistar}{\Phi^{\star}}
\newcommand{\Lstar}{L^{\star}}
\newcommand{\Mstar}{M^{\star}}
\newcommand{\SFRd}{\rho_{\rm SFR}}
\newcommand{\iyr}{yr$^{-1}$}
\newcommand{\vMpc}{Mpc$^{-3}$}

\newcommand{\fluxunit}{erg s$^{-1}$ cm$^{-2}$}

\title{Extinction Correction significantly influences the estimate of \Lya\ escape fraction}
\author{FANG~XIA AN\altaffilmark{1,2}, XIAN~ZHONG ZHENG\altaffilmark{1},  CAI-NA HAO\altaffilmark{3}, JIA-SHENG HUANG\altaffilmark{4}, AND XIAO-YANG XIA\altaffilmark{3}}

\altaffiltext{1}{Purple Mountain Observatory, China Academy of Sciences, Nanjing 210008, China; fangxiaan@pmo.ac.cn, xzzheng@pmo.ac.cn}
\altaffiltext{2}{University of Chinese Academy of Sciences, Beijing 100049, China; }
\altaffiltext{3}{Tianjin Astrophysics Center, Tianjin Normal University, Tianjin 300387, China}
\altaffiltext{4}{National Astronomical Observatories, Chinese Academy of Sciences, Beijing 100012, China}

%\altaffiltext{$\dagger$}{Based on observations obtained with WIRCam, a joint project of CFHT, Taiwan, Korea, Canada, France, and the Canada-France-Hawaii Telescope (CFHT) which is operated by the National Research Council (NRC) of Canada, the Institute National des Sciences de l'Univers of the Centre National de la Recherche Scientifique of France, and the University of Hawaii.}

\begin{abstract}
The \Lya\ escape fraction is a key measure to constrain the neutral state of the intergalactic medium and then to understand how the universe was fully reionized. We combine deep narrowband imaging data  from the custom-made filter NB393 and the $H_{2}S$1 filter centered at 2.14\,\mm\ to examine the \Lya\ emitters and \Ha\ emitters at the same redshift $z=2.24$. The combination of these two populations allows us to determine the \Lya\ escape fraction at $z=2.24$. Over an area of 383\,arcmin$^{2}$ in the Extended $Chandra$ Deep Field South (ECDFS), 124 \Lya\ emitters are detected down to NB393 = 26.4\,mag at the 5$\sigma$ level, and 56 \Ha\ emitters come from An et al. Of these, four have both \Lya\ and \Ha\ emissions (LAHAEs). We also collect the \Lya\ emitters and \Ha\ emitters at $z=2.24$ in the COSMOS field from the literature, and increase the number of LAHAEs to 15 in total. About one-third of them are AGNs. 
We measure the individual/volumetric \Lya\ escape fraction by comparing the observed \Lya\ luminosity/luminosity density to the extinction-corrected \Ha\ luminosity/luminosity density. We revisit the extinction correction for \Ha\ emitters using the Galactic extinction law with the color excess for nebular emission. We also adopt the Calzetti extinction law together with an identical color excess for stellar and nebular regions to explore how the uncertainties in extinction correction affect the estimate of individual and global \Lya\ escape fractions. In both cases, an anti-correlation between the \Lya\ escape fraction and dust attenuation is found among the LAHAEs, suggesting that dust absorption is responsible for the suppression of the escaping \Lya\ photons. However,  the estimated \Lya\ escape fraction of individual LAHAEs varies by up to $\sim 3$ percentage points between the two methods of extinction correction. We find the global \Lya\ escape fraction at $z=2.24$ to be ($3.7\pm1.4$)\% in the ECDFS. The variation in the color excess of the extinction causes a discrepancy of $\sim 1$ percentage point in the global \Lya\ escape fraction.

\end{abstract}

\section{INTRODUCTION}\label{s:introduction}
The \Lya\ emission line is a widely used probe for the physical properties of the interstellar medium (ISM) and intergalactic medium (IGM). 
One-third of the ionizing photons from hot stars could be reprocessed into the \Lya\ emission line through excitation and recombination of the atomic hydrogen gas \citep{Osterbrock62}, and  
thus are sufficiently bright to be detected in galaxies over a wide redshift range out to $z\sim 7$.
However, \Lya\ is often heavily attenuated by the dusty ISM around the young stars in star-forming galaxies (SFGs).
Compared with galaxies selected by the UV continuum or optical emission lines such as \Ha, \Lya\ emitters (LAEs) are 
usually recognized as galaxies in an early stage of formation, with smaller sizes, lower masses, lower metallicities, and younger ages \citep[e.g.,][]{Hayes10a,Cowie11,Nakajima13}.
The number density of LAEs increases from $z \sim 2$ to $z \sim 3$ \citep[e.g.,][]{Ciardullo12} and has little evolution at $z \sim$ 3-6 \citep[e.g.,][]{Ouchi08}, while the ultraviolet (UV) luminosity function (LF) of Lyman-break galaxies decreases rapidly during this period \citep{Bouwens15,Finkelstein15}, indicating that the LAEs play an important role in the assembly history of galaxies.
However, at higher redshift ($z > 6$), the number density of LAEs suddenly decreases because of the increasing fraction of neutral hydrogen (\HI) \citep[e.g.,][]{Ono10, Santos16}.

At $z \sim$ 2-3, the \Lya\ emission line can be observed in the optical window, and the amount of ionizing photons that escape from galaxies can be directly measured in the near-UV  \citep{Sandberg15, Matthee16b}.
%The redshift, $z\sim 2$, is the lowest redshift that LAEs can be observed in the optical window and, crucially, the amount of ionizing photons that escape from galaxies can be measured directly by stacking their near ultraviolet (UV) images \citep{Sandberg15, Matthee16}. 
The sizable sample of LAEs at this redshift range and the available multiwavelength observations allow us to investigate the physical properties of LAEs in detail, including stellar mass, dust attenuation, and star formation rate (SFR), and the correlations of these parameters with the escape fraction of \Lya\ photons \citep{Hayes10a, Atek14, Matthee16a}. By extrapolating these correlations to the high-$z$ universe, where the direct detection of ionizing photons is challenging or even practically impossible due to the increasing optical depth of the IGM with redshift, the role of SFGs in reionizing the early Universe can be investigated \citep{Hayes11, Bouwens16, Matthee16b}.

It has been suggested that LAEs significantly contribute to the cosmic reionization in the early universe \citep{Ono10,Nakajima13}. 
The shape of the \Lya\ line profile provides key constraints on the stage of cosmic reionization \citep{Kashikawa06, Dijkstra10}, and the gradual increase or sharp drop of the number density of LAEs reveals the changes in the ionization state of the IGM \citep[e.g.,][]{Malhotra06}.
Because of the resonant nature of the \Lya\ emission line, i.e., the \Lya\ photons are resonantly scattered by surrounding \HI,  
its escape fraction will hold key information on the physical properties and the environment conditions, such as the state of ionization and the distribution of ISM \citep[e.g.,][]{Neufeld91, Hansen06, Verhamme06, Schaerer11}. 
However, this resonant nature extends the escape path of \Lya\ photons and then enhances its absorption by dust.
In addition, the escape of \Lya\ photons strongly depends on the relative kinematics of the \HII\ and \HI\ regions, dust content, and geometry \citep{Giavalisco96, Kunth98, Mas03, Deharveng08, Hayes14}, making the determination of the escape fraction to be extremely complex and only empirically done.

Much effort has been dedicated to estimating the \Lya\ escape fraction in the local and the distant universe \citep[e.g.,][]{Deharveng08, Hayes10a, Blanc11, Atek14, Oteo15, Hagen16,Sobral16b} by comparing \Lya\ to other galactic emissions, like \Ha\ \citep{Hayes10a, Matthee16a}, \Hb\ \citep{Ciardullo14}, rest-frame UV \citep{Ouchi08, Konno16}, far-infrared\citep[FIR;][]{Wardlow14}, or X-ray \citep{Zheng12}, which indicate the young stars and star formation in a galaxy. 
Among these tracers, \Ha\ and \Hb\ are more direct probes of hot, young stars. 
The \Lya\ escape fraction can thus be determined by comparing the measurements of emissions from the same stellar population and without any empirical calibrations \citep{Hayes10a, Ciardullo14, Oteo15}. 
However, such measurements are still limited by a small survey area \cite[$\sim$ 56 arcmin$^{2}$;][]{Hayes10a}, limited galaxy sample size \citep{Ciardullo14}, or the lack of \Lya\ LF \citep{Oteo15}. 
Moreover, the measurements of both individual and volumetric \Lya\ escape fractions are very sensitive to the extinction correction of the \Ha\ sample. 
In our previous work \citep[][hereafter An14]{An14}, we employed the \cite{Calzetti00} extinction law and assumed the same color excess ($E(B-V)$) for nebular lines and the stellar continuum to correct the dust attenuation for \Ha\ emitters at $z = 2.24$, following the same method widely adopted in previous studies on high-$z$ galaxies. Recently, accumulating evidence has suggested that emission lines tend to be more attenuated than the stellar continuum in high-$z$ galaxies \citep[e.g.,][]{Forster09, Wuyts13, Reddy15}.
\cite{Wuyts13} found an extra extinction related to the \HII\ regions after calibrating \Ha-derived SFR with UV + IR-derived SFR. 
With high-quality near-IR (NIR) spectroscopic data, \cite{Reddy15} carried out a detailed study of dust attenuation of SFGs at $z \sim 2$ and found that the attenuation difference between ionized gas and stellar continuum strongly correlates with the SFR. 

In this work we present our measurements of the \Lya\ escape fraction.
We carry out a deep narrowband imaging survey through the custom-made NB393 filter in the Extended $Chandra$ Deep Field South (ECDFS) to search for \Lya\ emitters at the same redshift as the \Ha\ emitters presented in An14 .
With the multiwavelength data of the ECDFS, we identify the \Lya\ emitters and determine the observed \Lya\ LF. 
Combining the \Lya\ sample of \cite{Nilsson09} and \Ha\ sample of \cite{Sobral13}, we compare the properties of \Lya\ and \Ha\ emitters in the COSMOS field. 
We investigate the influence of extinction correction by using different extinction laws together with different assumptions of color excess for nebular lines.
%how the extinction correction influence the results %The \Lya\ escape fraction of individual shows an anti-correlation with the dust attenuation. {\bf The different assumptions of nebular color excess will dramatically change individual \Lya\ escape fraction and thus correlation between \Lya\ escape fraction and dust attenuation.}
%We determine the volumetric \Lya\ escape fraction by comparing the observed \Lya\ luminosity density to intrinsic \Ha\ luminosity density with the Case B assumption. {\bf The different assumptions of extinction correction will cause a discrepancy of 1 percentage point of global \Lya\ escape fraction at $z=2.24$.}
{\bf Section}~\ref{s:observation} presents the observations and data reduction. We describe how to measure the continuum and select \Lya\ emitters in {\bf Section}~\ref{s:sample}.
Our main results are shown in {\bf Section}~\ref{s:result} and we discuss and summarize our results in {\bf Section} \ref{s:discussion}. 
Throughout this paper we adopt a cosmology with [$\Omega_{\Lambda}$, $\Omega_M$, $h_{70}$] = [0.7, 0.3, 1.0]. The AB magnitude system \citep{Oke74} is used unless otherwise stated.

%\section{OBSERVATIONS AND SAMPLE SELECTION} \label{s:observation}
\section{OBSERVATIONS AND DATA REDUCTION} \label{s:observation}
\begin{figure}
\centering
\includegraphics[width=0.45\textwidth]{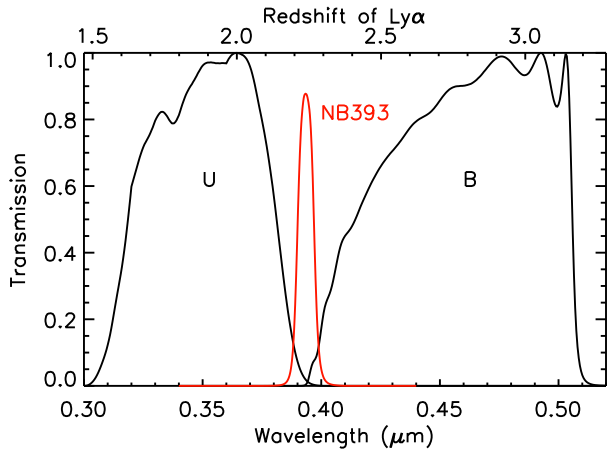}
\caption{Transmission curves. The red solid line shows the transmission curve of NB393 that is sensitive to \Lya\ emitters at $z$ = 2.24. The black lines are curves of the MUSYC $U$- and $B$-band.}
\label{f:tsc.eps}
\end{figure}

%\subsection{DUAL FILTER}
With Megacam on board the Magellan II telescopes \citep{Shectman03}, we obtained deep narrowband imaging data through the custom-made NB393 filter ($\lambda_\mathrm{c}$ = 3935.3$\,$\AA$, \Delta \lambda$ = 75.7\,\AA), which is designed to identify \Lya\ emitters at the same redshift as the \Ha\ emitters in our previous work An14. 
The NB393 observations were carried out under seeing conditions of $\sim$0$\farcs$6 with a total integration time of 11.75\,hr over three nights from 2011 December 27 to Dec 29 (PI: J.-S. Huang). In total, 47 exposures were taken, each with exposure time of 900\,s.
The pointing was slightly dithered to cover the small gaps between CCDs.
Broadband imaging data from the MUSYC survey are used to determine the continuum under the \Lya. 
We show the transmission curves of NB393 and MUSYC $U$- and $B$-band \citep{Gawiser06} in Figure~\ref{f:tsc.eps}.
The $U$-band science image from MUSYC is relatively shallow and only 45\% of NB393-detected sources have counterparts in the $U$-band. 
We construct SEDs from the $U, B, V, R$-band photometry for all NB393 sources and interpolate the continuum at 3935.3$\,$\AA.
Megacam's focal plane detector consists of 36 CCD42-90 CCDs manufactured by E2V, each with 2048 $\times$ 4608 pixels, and covers a large field of view 25$\arcmin \times 25\arcmin$ with a pixel scale of 0$\farcs$16\,pixel$^{-1}$ \citep{McLeod15}. 
The NB393 observations cover the survey area of 20$\arcmin \times 20\arcmin$ for \Ha\ emitters in ECDFS.
The narrowband data used in this analysis are part of a narrowband survey program (C.-N. Hao et al. 2017, in preparation).

We reduce the data following the Megacam data reduction procedure presented by Matthew~L.~N.~Ashby$\footnote{https://www.cfa.harvard.edu//$\sim$mashby/megacam/megacam
\_frames.html}$ and using the software IRAF together with the Megared package$\footnote{https://www.cfa.harvard.edu/$\sim$bmcleod/Megared}$. In the reduction procedure, bias subtraction is done in a standard way. About 20 twilight flat fields taken at the beginning and the end of each observing night are used to generate an average frame for flat fielding. Also, a mean sky background frame is created by combining individual exposures with objects masked. This mean sky background is scaled to match the background of each exposure image and removed in order to get rid of the changes in background due to the off-axis scattered light. The MUSYC catalog from \cite{Cardamone10} is used for astrometric calibration. The software tool Swarp \citep{Bertin02}  is used to mosaic 36 CCD images of each exposure into a single frame image, and co-add all exposures into the final mosaic image.

\begin{figure*}
\centering
\includegraphics[width=0.90\textwidth]{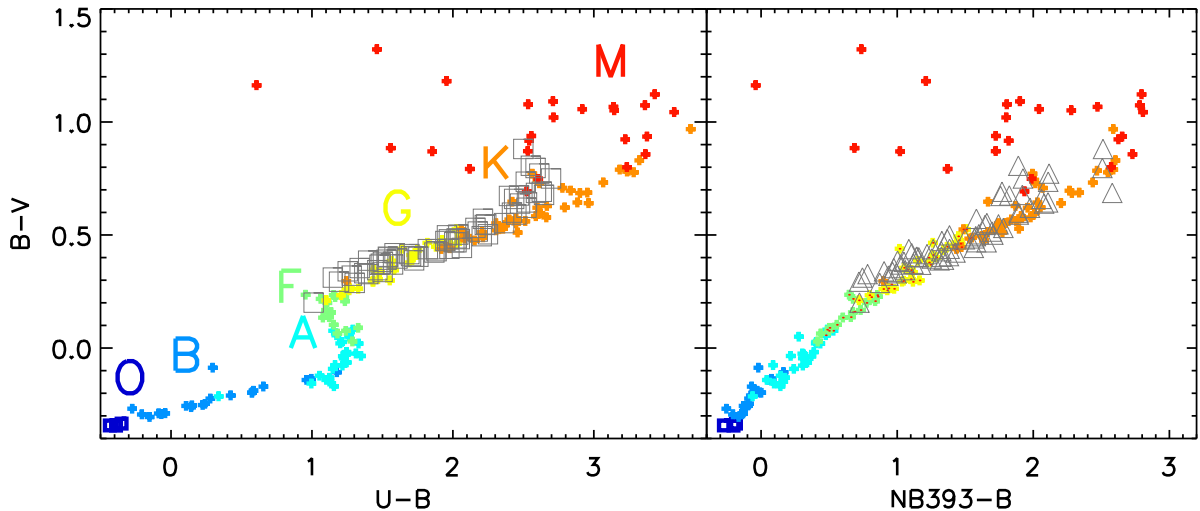}
\caption{Color-color diagrams to perform the photometry calibration of the NB393 images. {\bf Left:} comparison of $U-B$ and $B-V$ colors between the library of BPGS stars \citep{Gunn83} and observed stars in ECDFS.
$U-B$ and $B-V$ colors of different types of stars from BPGS are marked by different symbols of different colors. 
The gray squares are observed colors of 65 stars from the MUSYC catalog \citep{Cardamone10}. Good agreement between the two colors is shown in this plot; 
{\bf Right:} $NB393-B$ versus $B-V$. The observed colors of 65 stars with photometry-calibrated NB393 data are shown by gray triangles.}
\label{f:mdc.eps}
\end{figure*}

We perform photometric calibration for our narrowband imaging data by matching the observed colors of stars to model colors from the BPGS library \citep{Gunn83}. 
First, we identify stars from the MUSYC catalog \citep{Cardamone10} with SExtractor \citep{Bertin96} CLASS-STAR parameter $> 0.95$ and $U$-band total magnitude $17.5 \le mag\_{U} \le21.5$. %between 17.5 to 21.5$\,$mag. 
The left panel of Figure~\ref{f:mdc.eps} shows the aperture-matched $UBV$ colors of the total of 65 stars (gray squares) in good agreement with the model colors by convolving star spectra of BPGS with MUSYC $UBV$-band transmission curves. 
%It means that we can calibrate our narrowband data by matching the measured colors to the model colors.
We also extracted the empirical point spread functions (PSFs) from MUSYC and NB393 imaging data using these point sources and derived the correction for the aperture photometry with a diameter ($D$) of 2$\arcsec$.
The FWHM of the PSF is 0$\farcs$62 in the final NB393 science image. After that, we matched aperture-corrected NB393 fluxes to the MUSYC photometric system and calibrated the photometry for our narrowband data by matching the $U-$NB393 and $B-V$ colors of the above 65 stars to the model colors from BPGS as shown in the right panel of Figure~\ref{f:mdc.eps}.
%The left plot of Figure~\ref{f:mdc.eps} shows the good agreement between the photometry calibrated $U$NB393$V$ colors and model colors. 
The final reduced NB393 image covers a 626\,arcmin$^{2}$ area where the 5\,$\sigma$ depth within the $D = 1\farcs$6 aperture corresponds to 26.4\,mag for point sources.

\section{SAMPLE SELECTION} \label{s:sample}
\subsection{Source Detection and Continuum Measurement}
We use the software SEXtractor to detect sources in the NB393 image, and the detection of a source requires a minimum of 10 contiguous pixels above 2.0\,$\sigma$ of the background. 
The exposure map is taken as the weight image to reduce spurious detections in low signal-to-noise ratio (S/N) regions. 
In total, 34,530 sources are securely detected. We separate stars from detected sources through their NB393 $J K_{\rm s}$ colors and the flux ratio of different apertures and the SExtractor CLASS-STAR parameter. Firstly, we convolve the NB393 image with the $K_{\rm s}$-band PSF to match the spatial resolution of the CFHT/WIRCam $J K_{\rm s}$-band images \citep{Hsieh12}. We identify 832 stars with the separations of the NB393 $J K_{\rm s}$ colors and flux ratio of the apertures of diameter 1$\arcsec$ and 4$\arcsec$ ($>$ 0.23) and CLASS-STAR parameter $> 0.95$.
% and 832 stars are identified with SExtractor CLASS-STAR parameter $> 0.95$ and flux ratio of the apertures of diameter 1$\arcsec$ and 4$\arcsec$ $>$ 0.23 at which galaxies and stars are obviously separated.
Excluding stars, we obtain a catalog of 30,174 sources over the 626\,arcmin$^{2}$ area.

We then derive the continuum $C393$ between 3880 and 3990\,\AA\ for these objects by fitting their SEDs with $U,B,V,R$-band data from MUSYC. 
The aperture-matched ($D=1\farcs$6) colors are obtained using the same method described in \cite{Cardamone10}.
The default template set from EAZY \citep{Brammer08} is used to fit SEDs with a fixed redshift $z=2.24$. 
%The \Lya\ adsorption feature in the spectral templates are corrected for the estimate of continuum at \Lya.
To estimate the continuum at NB393, fixing the redshift of SEDs at $z=2.24$ is appropriate for LAEs. However, spectral features of \Lya\ absorption at $z=2.24$ may affect the continuum estimation for sources at other redshifts. We reduced this influence by interpolating the continuum from the continua next to the absorption.
Eleven objects are not resolved in the MUSYC images and excluded from our catalog. 
Finally, we obtain the SEDs of 30,163 sources. We then derive the model fluxes of NB393 and MUSYC $B$-band by convolving the SEDs with the respective transmission curves. The model fluxes of the continuum, $C393_{\rm SED}$, are matched to the MUSYC photometric system by multiplying with the ratio of $B_{\rm aper}$/$B_{\rm SED}$. Here, $B_{\rm aper}$ is the aperture-corrected $B$-band flux.

We convolve the NB393 image with the $B$-band PSF to match the spatial resolution of the MUSYC $B$-band image. We use the same aperture ($D=1\farcs$6) and the aperture correction in the $B$-band from \cite{Cardamone10} for the NB393 photometry to match its flux to the MUSYC photometry. Therefore, the color of $C393$ and $N393$ is matched indirectly. Figure~\ref{f:cdfs_mag_sele.eps} shows the color $C393-N393$ as a function of NB393 magnitude for the final 30,163 sources. 
The objects with MUSYC $B$-band magnitude $mag\_B > 26.3$\,mag (5\,$\sigma$ limit) are presented by the density map, and the solid points are sources with 27.0 (3$\sigma$ limit) $< mag\_B < 26.3$. 
In total 332 NB393-detected objects have $mag\_B < 27.0$, and we adopt the $B$-band 3\,$\sigma$ limit, $mag\_B$ = 27.0\,mag, as their continuum upper limit directly, shown with upward arrows in Figure~\ref{f:cdfs_mag_sele.eps}.

\subsection{Selection of Emitter Candidates}

\begin{figure}
\centering
\includegraphics[width=0.45\textwidth]{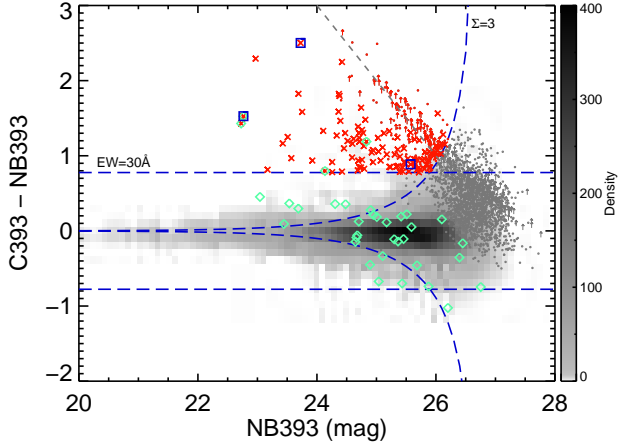}
\caption{Color-magnitude diagram for NB393-detected sources. 
The NB393 magnitude is the $MAG\_BEST$ given by SEXtractor, and $C393$ refers to the continuum magnitude derived from SEDs.
The gray scale image shows the number density of NB393-detected sources with secure $B$-band counterpart ($>5\,\sigma$). 
The solid points represent the NB393 sources with 3$\,\sigma < mag\_B < 5\,\sigma$. 
We use the $B$ band 3\,$\sigma$ limit as the continuum for the NB393 sources with $mag\_B < 3\,\sigma$ directly and we show their color lower limit with upward arrows.
The gray dashed line denotes continuum equal to the 3\,$\sigma$ limit in the $B$-band as a function of NB393 magnitude.
The blue dashed lines represent the limitation due to background noise at the $\Sigma=3$ level and the excess cut corresponding to the rest-frame $EW=30$\,\AA.
The red symbols show the 254 emission-line candidates with $\Sigma >$ 3 and $EW >$ 30\,\AA. Candidates with $mag\_B >$ 5\,$\sigma$ are marked with red crosses. 
There are 35 of the 56 \Ha\ emitters of An14 are detected in the NB393 observations and shown by light green diamonds. 
The blue squares mark the candidates with spec-$z =2.2$.}
\label{f:cdfs_mag_sele.eps}
\end{figure}

The method of selecting emitter candidates is the same as the one in An14. The narrowband excess is determined by a significance factor $\Sigma$ and a cut in the rest-frame equivalent width (EW) of an emission line represented by the dashed blue lines in Figure~\ref{f:cdfs_mag_sele.eps}. 
$\Sigma$ is the significance factor of the combined background noise of the broad and narrow bands as shown in Equation (1) in An14. In this work, the MUSYC $B$-band is used as the broadband.

Since the observed \Lya\ emission line will be heavily attenuated by dust and resonantly scattered by neutral \HI\ gas, the empirical cut for \Lya\ is lower than that for \Ha\ \citep[e.g.,][]{Nilsson09, Guaita10, Hayes10a}. 
In this work, we adopt 30\,\AA\ as a cut, which is estimated by the dispersion envelope as shown in Figure~\ref{f:cdfs_mag_sele.eps}.

Finally, we select 254 \Lya\ emission-line candidates with $\Sigma >$ 3 and $EW >$ 30\,\AA, marked by red symbols in Figure~\ref{f:cdfs_mag_sele.eps}. 
We visually examined the candidates in the NB393 and MUSYC images and found 13 of them to be close to a bright source in the MUSYC images and their continua are thus difficult to estimate correctly. 
Therefore, we removed these sources for simplificity. 
The possible emission lines contaminating \Lya\ are \OII$\lambda$3727 at $z=0.06$ and the doublet line of \CIV$\lambda\lambda$1548,1550\,\AA\ at $z=1.54$ from active galactic nuclei (AGNs). 
We point out that the \OII\ emitters would have a comparatively tiny cosmic volume and are too few to contaminate our sample. 
On the other hand, the local universe \OII\ emitters would have a rather extended appearance and stand out clearly in our catalog. Our visual examination confirms that none of our candidates is an \OII\ emitter. 
We cross-correlate our sample with the 4 Ms Chandra observations of the ECDFS \citep{Xue11}. 
Six of the 241 sample galaxies are found to be X-ray sources (ID=29, 101, 257, 440, 473, 720 in the Chandra catalog) and are all classified as AGNs with $L_\mathrm{0.5-8\,keV} \ge 3\times 10^{42}$\,erg\,s$^{-1}$. 
However, previous works have revealed that the sample of \Lya\ emitters contains a comparable fraction of AGNs at this redshift \citep{Nilsson11}. 
We cannot determine whether these six sources are LAEs at $z=2.24$ or \CIV$\lambda\lambda$1548,1550\,\AA\ at $z=1.54$. 
Nevertheless, we keep the X-ray sources in our sample and come back to this in Section ~\ref{s:global fraction}. 

\section{The \Lya\ Escape Fraction at $z$ = 2.24} \label{s:result}

\subsection{Observed \Lya\ and \Ha\ luminosities} \label{s:luminosities}
We aim to study the connections between \Lya\ and \Ha\ emitters at the same cosmic time. 
The NB393 observations overlap with the survey area of the 56 \Ha\ emitters from An14, yielding 124 of 241 LAEs in the overlapping region of 383 arcmin$^{2}$. 
Four of the 124 \Lya\ emitters are also \Ha\ emitters. 
In turn, 35 of the 56 \Ha\ emitters are detected in NB393, which are marked by light green diamonds in Figure~\ref{f:cdfs_mag_sele.eps}.
The reasons that two kinds of emitters are seldom overlap have been discussed in several studies \citep[e.g.,][]{Hayes10a, Oteo15}.
In short, the two lines have different sensitivity to dust absorption, and the \Lya\ emission line can be resonantly scattered, resulting in the narrowband-selected \Lya\ and \Ha\ emitters coming from different populations.
\Lya\ emitters are relatively less massive and contain less dust than \Ha\ emitters. We abbreviate these common emitters as \Lya\ and \Ha\ emissions (LAHAEs).

\begin{figure}
\centering
\includegraphics[width=0.45\textwidth]{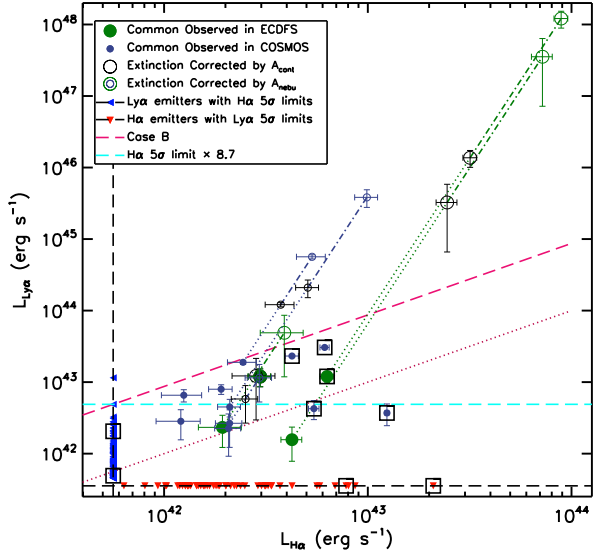}
\caption{Luminosity comparison of \Ha\ and \Lya\ emitters at $z$ =2.24 in the ECDFS. The red and blue triangles represent the \Ha\ emitters without \Lya\ emission-line detection, and \Lya\ emitters without \Ha\ detection respectively. 
The green solid points are the four LAHAEs with 1$\sigma$ photometric uncertainties. 
The black open circles show the \Ha\ and \Lya\ luminosities corrected for dust attenuation derived from the continuum ($A\rm_{cont}$). The open colored circles represent the nebular extinction-corrected \Ha\ and \Lya\ luminosities. The observed data points, $A\rm_{cont}$-corrected luminosities and $A\rm_{nebu}$-corrected luminosities, are connected by the dotted and dotted-dashed lines, respectively. The intrinsic \Lya/\Ha\ luminosity ratio of 8.7 for case B recombination is shown by the magenta dashed line. 
Eleven LAHAEs from the COSMOS field are shown by small blue solid points. The three dust-corrected LAHAEs in the COSMOS field are shown by small open circles. The black squares mark AGNs in both samples.}
\label{f:luminosities.eps}
\end{figure}

The MAG\_BEST from SExtractor is used as the total magnitude to determine the \Lya\ luminosities. Contributions from the continuum are subtracted using $C393-NB393$ colors.
We present the observed \Lya\ and \Ha\ luminosities in the overlapping region of the ECDFS at $z$ = 2.24 in Figure~\ref{f:luminosities.eps}. 
The \Ha\ emitters without \Lya\ emission-line detection are presented as red triangles, and the \Lya\ emitters without \Ha\ detection are shown as blue triangles. 
The green solid points are the four LAHAEs with error bars to show 1$\sigma$ photometric uncertainties. 

\subsection{Extinction correction} \label{s:corrected}

As described earlier, extinction correction for \Ha\ emission is crucial in estimating the \Lya\ escape fraction.  
For our 56 \Ha\ emitters, we have fitted their SEDs by using 12 band imaging data from MUSYC $U$ \citep{Gawiser06} to CFHT/WIRCam $K{\rm_s}$ \citep{Hsieh12} and narrow H$_2$S1-band data.
Details of the photometry and aperture-matched colors can be found in section 3.2 in An14. In An14, the dust attenuation was estimated by SED fitting, and extinction correction for the 56 \Ha\ emitters was made using 
the Calzetti extinction law \citep{Calzetti00} with the assumption of an identical color excess for stellar and nebular regions ($E(B-V)\rm_{cont}$ = $E(B-V)\rm_{nebu}$). However, recent studies show that the nebular emission lines suffer an increasing degree of obscuration relative to the continuum, and the Galactic extinction curve is favored for the lines \citep[e.g.,][]{Forster09, Wuyts13, Reddy15}. 
We repeat the extinction correction for the \Ha\ emitters using the \cite{Cardelli89} Galactic extinction law with R$\rm_{v}$ = 3.1 and the prescription of nebular color excess in \cite{Wuyts13}.
%Therefore, we modify the extinction correction for the \Ha\ sample in An14 by employing the Cardelli Galactic extinction law with R$\rm_{v}$ = 3.1 and assuming the prescription of color excess between nebula and continuum in \cite{Wuyts13}.  
To investigate how extinction corrections influence both the individual and global \Lya\ escape fractions presented in Section~\ref{s:individual fraction} and Section~\ref{s:global fraction}, we also use the Calzetti extinction law and assume the same color excess for the ionized gas and stellar continuum as widely adopted in the literature.

Replacing the Calzetti extinction law with the Cardelli Galactic extinction law only causes a discrepancy of $<$ 0.01\,mag in dust attenuation for the \Ha\ continuum, but a discrepancy of 0.58\,mag for the \Lya\ continuum. Therefore, the change of extinction laws affects neither individual nor global \Lya\ escape fraction. In this work, we firstly correct \Ha\ luminosities for the dust attenuation derived from the continuum ($A\rm_{cont}$). Then we employ the prescription given in \cite{Wuyts13},
\begin{equation} \label{e:extra}
\begin{split}
A\rm_{extra} =0.9\,A\rm_{cont}-0.15\,A^{2}\rm_{cont}
\end{split}
\end{equation}
to correct the extra extinction for \Ha\ emitters. For \Lya\ emitters, we use the  $E(B-V)\rm_{nebu}$ derived from nebular dust attenuation of \Ha\ emitters ($A\rm_{nebu}$ = $A\rm_{cont}$ + $A\rm_{extra}$) and the \cite{Cardelli89} extinction law to correct the dust extinction.
We connect the dust-corrected data points (open circles) and the observed \Lya\ and \Ha\ luminosity data points with dotted and dotted-dashed lines in Figure~\ref{f:luminosities.eps}. The black circles represent the luminosities corrected by $A\rm_{cont}$ and the color circles show the data corrected by $A\rm_{nebu}$. It is obvious that the assumption of nebular color excess will largely influence the extinction correction and our main results. 
%where $A\rm_{nebu}$ =  $A\rm_{cont}$ + $A\rm_{extra}$. }

The magenta dashed line in Figure~\ref{f:luminosities.eps} shows the intrinsic \Lya/\Ha\ luminosity ratio of 8.7 for the case B recombination. 
It can be seen that two of the LAHAEs have \Lya/\Ha\ ratio $>$ 8.7 even if only $A\rm_{cont}$ is used to correct extinction. Additional extinction correction for LAHAEs increases their number from two to three and makes the extinction-corrected \Lya/\Ha\ ratio deviate further away from the theoretical ratio in case B.
It suggests that the assumption of case B recombination may not be a good approximation at high redshift \citep{Song14}. 
%It may also possibly attributed to dust attenuation curve at high redshift different from that of local starburst galaxies \citep{Buat12, Zeimann15}. 
It might be partially attributed to the dust attenuation curve in high-$z$ galaxies differing from the adopted ones for local galaxies \citep{Buat12, Zeimann15} or/and the large uncertainties in the correlation between nebular and stellar color excess \citep{Wuyts13, Reddy15}.

%or the Calzetti extinction law from local starburst galaxies might not be suitable for high redshift emission-line galaxies directly. 

\begin{figure}
\centering
\includegraphics[width=0.45\textwidth]{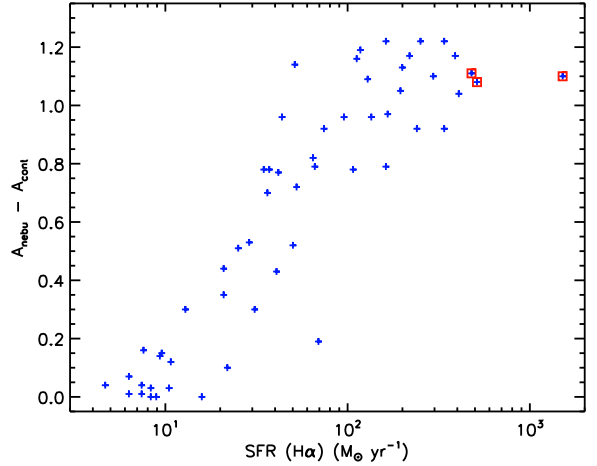}
\caption{Difference between nebular and stellar dust attenuation as a function of SFR for 56 \Ha\ emitters at $z=2.24$. \cite{Cardelli89} Galactic extinction curve is assumed. $A\rm_{nebu}=A\rm_{cont}+A\rm_{extra}$, where $A\rm_{extra}$ is derived by using the prescription in \cite{Wuyts13}. Three X-ray sources are marked by red squares.}
\label{f:extinction.eps}
\end{figure} 

We also plot eleven LAHAEs (small blue solid points in Figure~\ref{f:luminosities.eps}) in the COSMOS field where the \Lya\ emitter sample is from \cite{Nilsson09} and the \Ha\ emitter sample is from HiZELS survey \citep{Sobral13}. 
\cite{Nilsson11} performed a detailed analysis and estimated dust extinction for 58 \Lya\ emitters. Three of the eleven LAHAEs are also in the 58 sources. We thus correct dust attenuation for their \Lya\ and \Ha\ luminosities. 
From Figure~\ref{f:luminosities.eps} we see that two of the three LAHAEs also have their intrinsic \Lya/\Ha\ ratio larger than the theoretical estimate with case B recombination.

Figure~\ref{f:extinction.eps} shows the difference between the dust attenuation of recombination line photons and continuum photons at 6563\,\AA\ as a function of SFR. We calculate the SFR from \Ha\ luminosity using the formula given in \cite{Kennicutt12}. A trend is clearly seen with the difference of dust attenuation between ionized gas and stellar continuum increasing with SFR. This is consistent with the result of \cite{Reddy15} based on a detailed study of dust attenuation in $z\sim2$ SFGs using NIR spectroscopic data.

For the four LAHAEs in our sample, one is an X-ray source with absorption-corrected rest-frame 0.5-8\,keV luminosity of $2.60\times 10^{43}$ erg s$^{-1}$ and is classified as an AGN \citep{Xue11}. 
For the LAHAEs in the COSMOS field, 4 of 11 are identified as AGNs \citep{Nilsson11}. 
The fraction of AGNs in LAHAEs is roughly one-thrid for the ECDFS and COSMOS samples.

\begin{figure}
\centering
\includegraphics[width=0.45\textwidth]{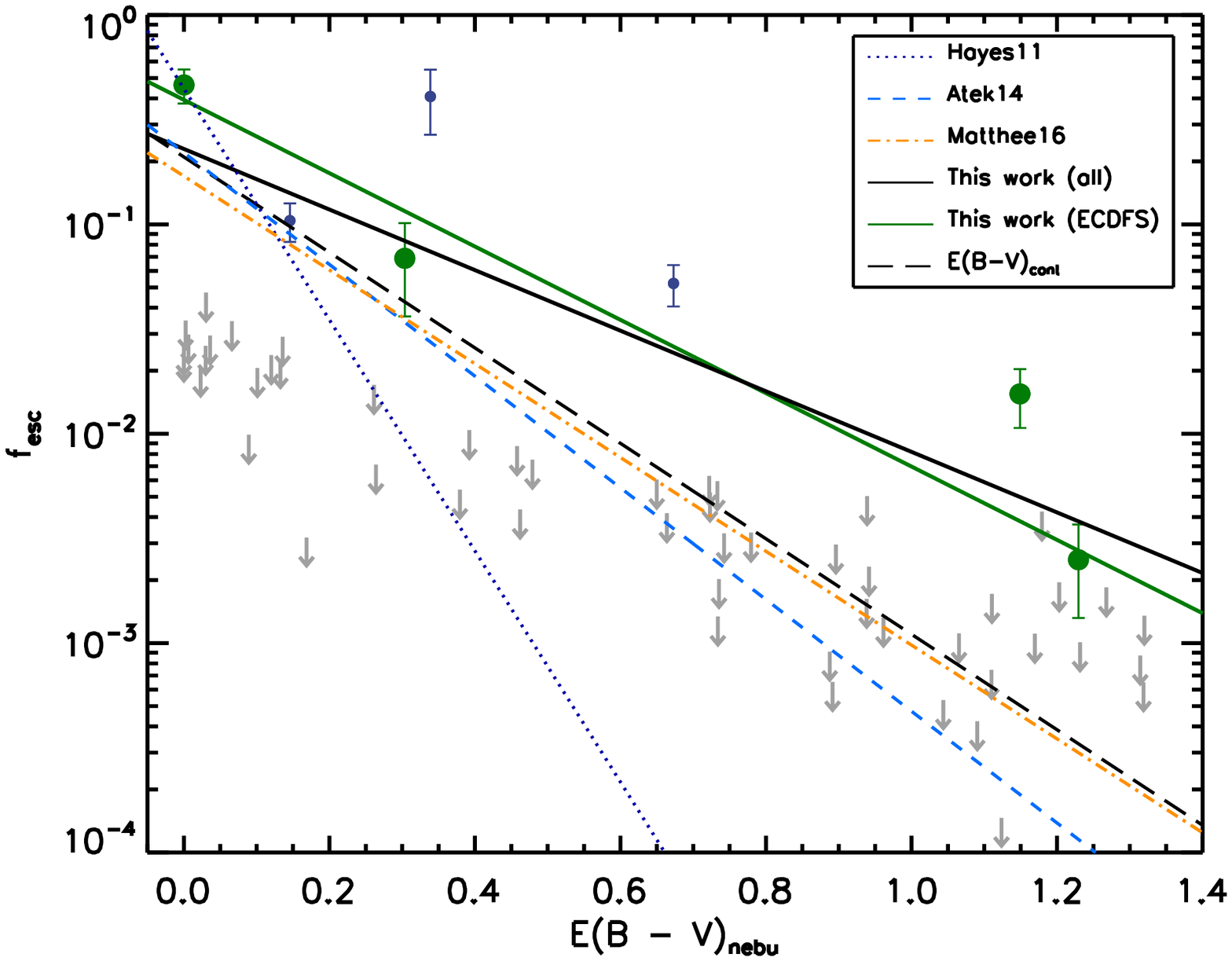}
\caption{\Lya\ escape fraction as a function of nebular $E(B-V)$ by employing the \cite{Cardelli89} Galactic extinction law. 
Four LAHAEs in the ECDFS are shown with green solid points. 
Arrows represent the upper limit of the \Lya\ escape fraction for 52/56 \Ha\ emitters without secure \Lya\ detections.
For 11 LAHAEs in the COSMOS field, SED-derived dust attenuation is available for three of them. 
We estimated the escape fraction of \Lya\ photons of the three sources and mark them with small blue points. 
The black solid line is the best-fit correction between $E(B-V)\rm_{nebu}$ and $f_{\rm esp}$ for all seven LAHAEs. The green solid line is the relation fitted by the four LAHAEs in the ECDFS. The results of previous woks are plotted for comparison. We also fit the correlation of $f_{\rm esp}-E(B-V)\rm_{cont}$ (black dashed line) to see how different assumptions of the nebular color excess change the results.}
\label{f:ind_esp.eps}
\end{figure}

\subsection{\Lya\ escape fraction in Individual galaxies} \label{s:individual fraction}
For the sample of 56 \Ha\ emitters, we individually correct their dust attenuation.
With the assumption of negligible optical depth in all but the Lyman series, the assumption of case B recombination \cite{Baker38} can properly reproduce observations of hydrogen emission from ISM.
We adopt the theoretical value of $L_{\rm {Ly\alpha}}/L_{\rm {H\alpha}}$ = 8.7 as the standard \Lya/\Ha\ line ratio. Then we estimate the \Lya\ escape fraction for individual galaxies using: 
\begin{equation} \label{e:schechter}
\begin{split}
f_{\rm esp} =L_{\mathrm{obs}}(\mathrm{Ly\alpha})/(8.7 \times L_{\mathrm{int}}(\mathrm{H\alpha})),
\end{split}
\end{equation}
where $L_{\mathrm{obs}}$(Ly$\alpha$) is the observed \Lya\ luminosity and  $L_{\mathrm{int}}$(H$\alpha$) is the dust-corrected \Ha\ luminosity. 
We present the \Lya\ escape fraction as a function of nebular dust extinction for the 56 \Ha\ emitters in Figure~\ref{f:ind_esp.eps}.
Since 52/56 \Ha\ emitters do not have secure \Lya\ detections, we use the 5\,$\sigma$ limit of \Lya\ luminosity, log $(L_{\mathrm{Ly\alpha}})$ = 43.55, to derive the upper limit of the \Lya\ escape fraction for these sources. 
The four LAHAEs are shown as green solid points and the three dust-corrected LAHAEs from the COSMOS field are also plotted in Figure~\ref{f:ind_esp.eps} as small blue points. 
It is obvious that dust prevents the escape of \Lya\ photons. Thus $f_{\rm esp}$ decreases as the dust attenuation increases, as shown in Figure~\ref{f:ind_esp.eps}. 
The correlation between $f_{\rm esp}$ and $E(B-V)\rm_{nebu}$ can be described by the function:
\begin{equation} \label{e:extinction}
\begin{split}
f_{\rm esp} =C_{\mathrm{Ly\alpha}} \times 10^{-0.4\,E(B-V)\,k_{\mathrm{Ly\alpha}}},
\end{split}
\end{equation}
where $C_{\mathrm{Ly\alpha}}$ is a constant coefficient and $k_{\mathrm{Ly\alpha}}$ is the extinction coefficient \citep{Hayes11, Atek14, Matthee16a}. By minimizing the $\chi^{2}$ for all data points, we find that the best-fit parameters are $C_{\mathrm{Ly\alpha}}=0.22 \pm 0.04$ and $k_{\mathrm{Ly\alpha}}=3.62\pm0.32$. We also fit the correlation only using the four LAHAEs in the ECDFS, yielding the best-fit parameters of $C_{\mathrm{Ly\alpha}}=0.39 \pm 0.08$ and $k_{\mathrm{Ly\alpha}}=4.38\pm0.41$ as shown by the green solid line in Figure~\ref{f:ind_esp.eps}. We plot in Figure~\ref{f:ind_esp.eps} the results from previous works  for comparison. We point out that \cite{Matthee16a} also used the sample from the HiZELS survey \citep{Sobral13} but with a new \Lya\ sample. The decrease of \Lya\ escape fraction with extinction is slightly steeper and higher than that given in \cite{Matthee16a} and \cite{Atek14}.
However, we notice that \cite{Matthee16a} used the Calzetti extinction law with an assumption of $E(B-V)\rm_{nebu}$ = $E(B-V)\rm_{cont}$. 
Therefore, we also fit the $f_{\rm esp}-E(B-V)$ relation by using $A\rm_{cont}$-corrected \Ha\ emitters and show the results with the black long-dashed line in Figure~\ref{f:ind_esp.eps}. Clearly, a good agreement is obtained then, implying that the assumption of nebular color excess is key to fitting the $f_{\rm esp}$-$E(B-V)$ correlation. 
\cite{Atek14} employed the gas-phase dust extinction and Cardelli extinction law as we did but for a lower-redshift sample at $z=0.3$. We argue that the discrepancy in the $f_{\rm esp}-E(B-V)$ relation between our work and \cite{Atek14} is probably caused by the different properties of SFGs at high-$z$, like having a much higher frequency of clumpy and irregular structures \citep[e.g.,][and references therein]{Shapley11}. As for the significantly steeper slope of the $f_{\rm esp}-E(B-V)$ relation in \cite{Hayes11}, it may be caused by missing dusty galaxies as explained in \cite{Oteo15} and \cite{Matthee16a}.
%we agree the explanation of missing dusty galaxies in this work as discussed in \cite {Oteo15} and \cite{Matthee16a}.}

%The black solid line describes the best fit to $f_{\rm esp}$ - $E(B-V)\rm_{nebu}$ correlation based on seven LAHAEs.}
%The dashed line describes the best fit to $f_{\rm esp}$ - $A_{V}$ correlation based on seven LAHAEs, having a slope of $- 0.58$. 
We list the coordinates, observed \Ha\ luminosities, dust attenuation ($A_{V}$), nebular extinction-corrected \Ha\ luminosities, the NB393 magnitudes, the SED-derived continuum, the \Lya\ luminosities, and the \Lya\ escape fractions of our 56 \Ha\ emitters in Table~\ref{table1}. 
It can be seen that $A_{V}$ for these LAHAEs range from zero to $\sim$ 2.5. 
Although the nearly dust-free LAHAEs do have larger \Lya\ escape fraction, their $f_{\rm esp}$ are still smaller than 50\%. 
Therefore, the lack of dust appears does not appear to be a requirement for the escape of \Lya\ photons. The resonant scatter on the neutral \HI\ gas can contribute to more than half of the suppression of \Lya\ escape in some cases. 
We notice that the different assumptions of nebular color excess will cause a discrepancy of up to $\sim 3$ percentage points in individual \Lya\ escape fraction estimation.
Besides the four LAHAEs, the \Lya\ escape fraction of \Ha\ emitters is the upper limit as described above. 
The NB393 magnitudes and the SED-derived continuum are only available for 35/56 \Ha\ emitters that have NB393 detections. 
We mark the \Lya\ luminosity as "$-99.99$" for the sources that have negative $NB393 - C393$ colors. The other properties, such as stellar mass and morphology, for the 56 \Ha\ emitters can be found in table 2 of An14. 

%\onecolumngrid
\begin{table*}[!h]
%\tablewidth{\textwidth}
\tabcolsep=11.0pt
\begin{center}
\caption{\label{table1} Individual Ly$\alpha$ Escape Fraction of the H$\alpha$ Sample}
\tiny 
\begin{tabular}{cccccccccc}
\\[0.3mm]
\hline \\[-2.3mm]
\hline \\[-2.1mm]
	ID & R.A.  & Decl. &log~$L_{\rm obs}$(${\rm H\alpha}$) &$A_{V}$  &log~$L_{\rm cor}$(${\rm H\alpha}$)\tablenotemark{a} & Mag\_NB393  & Mag\_C393  &log~$L_{\rm obs}$(${\rm Ly\alpha}$)  &  f$_{esp}$  \\
	& (J2000) & (J2000) & (erg~s$^{-1}$) & (mag) & (erg~s$^{-1}$)   & (AB)&    (AB)  & (erg~s$^{-1}$) & $(\%)$ \\
\hline \\[-2mm]
    1      &53.254681     &$-$27.890129  &42.57 &1.30 &43.30   & 24.70$\pm$0.04   & 24.83$\pm$  0.15    & 41.46 &$<$0.2\tablenotemark{b}  \\
    2      &53.060623     &$-$27.882366  &42.63 &2.35 &43.86    & 24.83$\pm$0.03   & 26.02$\pm$  0.22    & 42.20 & 0.3       \\ 
    3      &53.195805     &$-$27.877535  &42.27 &2.08 &43.38    &... \tablenotemark{c}              &... &...    &$<$0.2      \\ 
    4      &53.109905     &$-$27.870167  &42.08 &2.44 &43.34    &...               &...                  &...    &$<$  0.2  \\ 
    5      &53.208305     &$-$27.851402  &42.51 &0.79 &42.97    & 23.04$\pm$0.01   & 23.50$\pm$  0.03    & 42.62 &$<$  0.4  \\ 
    6      &53.054199     &$-$27.821192  &42.22 &0.00 &42.22    & 24.48$\pm$0.03   & 24.84$\pm$  0.13    & 41.96 &$<$  2.4  \\  
    7      &53.106880     &$-$27.817715  &42.12 &0.44 &42.38    & 25.51$\pm$0.04   & 25.73$\pm$  0.11    & 41.37 &$<$  1.7  \\ 
    8      &53.110756     &$-$27.797260  &41.97 &1.71 &42.91    &...               &...                  &...    &$<$  0.5  \\ 
    9      &53.099648     &$-$27.794247  &42.75 &1.30 &43.48    & 25.88$\pm$0.07   & 25.14$\pm$  0.16    &$-$99.99 \tablenotemark{d} &$<$  0.1 \\ 
   10      &53.099121     &$-$27.786957  &42.26 &0.82 &42.73    &...               &...                  &...       &$<$  0.8  \\ 
   11      &53.117970     &$-$27.786526  &42.30 &1.72 &43.25    & 26.40$\pm$0.06   & 26.04$\pm$  0.13    &$-$99.99  &$<$  0.2  \\
   12      &53.106899     &$-$27.786377  &42.21 &0.78 &42.67    & 25.68$\pm$0.05   & 25.22$\pm$  0.10    &$-$99.99  &$<$  0.8  \\
   13      &53.112999     &$-$27.778648  &42.47 &0.00 &42.47    & 22.76$\pm$0.01   & 24.29$\pm$  0.08    & 43.08    &46.4      \\ 
   14      &53.112816     &$-$27.776285  &42.19 &0.00 &42.19    &...                     &...            &...       &$<$  2.6      \\ 
   15\tablenotemark{*}      &53.121212  &$-$27.774660 &42.89 &2.03 &43.98   & 24.89$\pm$0.03    & 24.44$\pm$0.06  &$-$99.99 &$<$  0.0 \\ 
   16\tablenotemark{*}      &53.131065  &$-$27.773066 &43.32 &2.11 &44.45  & 24.31$\pm$0.02    & 24.66$\pm$0.08  &42.03 &$<$  0.0 \\ 
   17      &53.154442     &$-$27.771420  &42.84 &1.93 &43.88    &  26.20$\pm$0.06   &  25.18$\pm$  0.09  &$-$99.99  &$<$  0.1     \\ 
   18      &53.139744     &$-$27.763180  &42.16 &2.56 &43.48    & ...               & ...                &...       &$<$  0.1   \\ 
   19      &53.121334     &$-$27.755846  &42.63 &2.08 &43.74    & ...               & ...                &...       &$<$  0.1   \\ 
   20      &53.168320     &$-$27.748516  &42.48 &2.56 &43.80    & ...               & ...                &...       &$<$  0.1   \\ 
   21      &53.144493     &$-$27.728069  &42.33 &1.16 &42.99    & ...               & ...                &...       &$<$  0.4   \\
   22      &53.174511     &$-$27.725378  &42.50 &0.64 &42.88    &  25.04$\pm$0.05   &  24.37$\pm$  0.10  &$-$99.99  &$<$  0.5     \\ 
   23      &53.018269     &$-$27.725254  &42.27 &0.04 &42.29    &  23.45$\pm$0.01   &  23.54$\pm$  0.03  & 41.84    &$<$  2.1   \\ 
   24      &53.034084     &$-$27.711220  &42.50 &1.98 &43.56    & ...               & ...                &...       &$<$  0.1   \\ 
   25      &53.159687     &$-$27.663280  &42.16 &0.05 &42.19    &  25.30$\pm$0.03   &  25.19$\pm$  0.09  &$-$99.99  &$<$  2.6     \\  
   26      &53.107876     &$-$27.710629  &41.91 &0.05 &41.94    &  25.01$\pm$0.03   &  25.19$\pm$  0.07  & 41.50    &$<$  4.7   \\ 
   27      &53.078926     &$-$27.676744  &42.12 &0.20 &42.24    &  25.46$\pm$0.03   &  25.35$\pm$  0.09  &$-$99.99  &$<$  2.4     \\ 
   28      &53.175816     &$-$27.972818  &42.18 &1.14 &42.83    &  26.10$\pm$0.05   &  26.25$\pm$  0.15  & 40.99    &$<$  0.6   \\
   29      &53.183155     &$-$27.972805  &42.10 &0.06 &42.14    &  25.59$\pm$0.04   &  25.64$\pm$  0.13  & 40.74    &$<$  2.9   \\ 
   30      &53.207333     &$-$27.967606  &41.80 &2.23 &42.98    & ...               & ...                &...       &$<$  0.4   \\ 
   31      &53.251198     &$-$27.946007  &42.13 &0.01 &42.14    &  24.65$\pm$0.04   &  24.50$\pm$  0.11  &$-$99.99  &$<$  3.0     \\ 
   32      &53.296516     &$-$27.920425  &42.90 &1.61 &43.80    &  24.66$\pm$0.03   &  24.58$\pm$  0.08  &$-$99.99  &$<$  0.1     \\ 
   33      &53.053555     &$-$27.917978  &42.40 &2.21 &43.57    &  25.10$\pm$0.05   &  24.77$\pm$  0.16  &$-$99.99  &$<$  0.1     \\ 
   34      &52.938030     &$-$27.878836  &42.49 &0.44 &42.76    &  24.68$\pm$0.03   &  24.62$\pm$  0.08  &$-$99.99  &$<$  0.7     \\ 
   35      &52.954445     &$-$27.863352  &42.38 &2.35 &43.61    & ...               & ...                &...       &$<$  0.1   \\ 
   36      &52.984840     &$-$27.857210  &42.24 &1.62 &43.14    & ...               & ...                &...       &$<$  0.3   \\ 
   37      &53.259026     &$-$27.822220  &42.12 &2.29 &43.32    & ...               & ...                &...       &$<$  0.2   \\ 
   38      &52.965557     &$-$27.820986  &42.17 &1.28 &42.89    &  26.75$\pm$0.09   &  26.01$\pm$  0.13  &$-$99.99  &$<$  0.5     \\ 
   39      &52.964603     &$-$27.820471  &42.12 &0.22 &42.25    & ...               & ...                &...       &$<$  2.3   \\ 
   40      &52.985733     &$-$27.809784  &42.30 &1.39 &43.08    & ...               & ...                &...       &$<$  0.3   \\ 
   41      &53.020000     &$-$27.803131  &42.46 &1.71 &43.40    & ...               & ...                &...       &$<$  0.2   \\ 
   42      &53.254181     &$-$27.786942  &42.01 &0.22 &42.15    &  25.17$\pm$0.03   &  25.28$\pm$  0.16  & 41.24    &$<$  2.9   \\ 
   43\tablenotemark{*}    &52.961563     &$-$27.784283  &42.80  &2.16 &43.95 & 22.73$\pm$0.01     & 24.16$\pm$0.04   &43.08  & 1.6  \\   
   44      &52.972492     &$-$27.778952  &42.10 &1.30  &42.84  & ...               & ...                & ...       &$<$  0.6   \\        
   45      &53.284180     &$-$27.759590  &42.07 &0.00  &42.07  &  25.42$\pm$0.03   &  25.61$\pm$  0.16  &  41.35    &$<$  3.5   \\ 
   46      &53.264011     &$-$27.756128  &42.34 &1.31  &43.09  & ...               & ...                & ...       &$<$  0.3   \\ 
   47      &53.246250     &$-$27.753157  &42.01 &0.11  &42.07  &  24.90$\pm$0.03   &  25.18$\pm$  0.09  &  41.70    &$<$  3.4   \\ 
   48      &53.272289     &$-$27.745440  &42.36 &2.55  &43.67  & ...               & ...                & ...       &$<$  0.1   \\
   49      &53.280537     &$-$27.735306  &42.20 &0.66  &42.59  &  24.95$\pm$0.03   &  25.16$\pm$  0.06  &  41.58    &$<$  1.0   \\ 
   50      &53.257740     &$-$27.725210  &42.53 &0.15  &42.61  &  25.36$\pm$0.05   &  25.22$\pm$  0.21  & $-$99.99  &$<$  1.0  \\ 
   51      &53.270630     &$-$27.724007  &42.53 &1.76  &43.49  &  23.69$\pm$0.02   &  23.99$\pm$  0.05  &  42.21    &$<$  0.1   \\ 
   52      &53.199146     &$-$27.675362  &42.76 &1.60  &43.65  &  25.43$\pm$0.04   &  24.73$\pm$  0.08  & $-$99.99  &$<$  0.1  \\ 
   53      &52.960171     &$-$27.692127  &42.20 &0.17  &42.30  & ...               & ...                & ...       &$<$  2.1   \\ 
   54      &53.276863     &$-$27.693634  &42.09 &1.28  &42.81  &  26.45$\pm$0.05   &  26.29$\pm$  0.16  & $-$99.99  &$<$  0.6  \\ 
   55      &53.036934     &$-$27.648577  &42.29 &0.51  &42.59  &  24.14$\pm$0.02   &  24.93$\pm$  0.08  &  42.37    &     6.9   \\ 
   56      &53.146427     &$-$27.658226  &42.94 &0.28  &43.11  &  23.54$\pm$0.01   &  23.90$\pm$  0.04  &  42.35    &$<$  0.3   \\ 
\hline

\end{tabular}
\\[1.5mm]
{$^{a}$Nebular dust extinction-corrected H$\alpha$ luminosities; $^{b}$ $"<"$ means that the Ly$\alpha$ escape fraction is the upper limit; \\
$^{c}$H$\alpha$ emitters without NB393-band detection; $^{d}$H$\alpha$ emitters have negative $NB393-C393$ color; $^{*}$Three X-ray sources.}
%\tablenotetext{*}
\end{center}

\end{table*}
%\twocolumngrid

\begin{figure}
\centering
\includegraphics[width=0.45\textwidth]{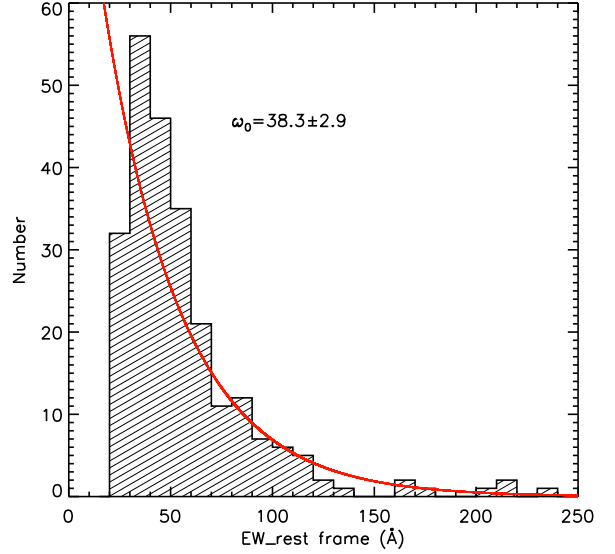}
\caption{Distribution of rest-frame EWs of \Lya\ emitters at $z$ = 2.24. 
The red line shows the best-fit exponential function with $\omega_{0}$ = 38.3 $\pm$ 2.9\,\AA.}
\label{f:ew.eps}
\end{figure} 

\subsection{The volumetric \Lya\ escape fraction at $z$ = 2.24} \label{s:global fraction}
The other main purpose of this work is to estimate the global \Lya\ escape fraction at $z$ = 2.24. 
We determine the \Lya\ LF using our 235 \Lya\ emission-line galaxies. 
Six X-ray sources are excluded to avoid AGN contamination as we did for \Ha\ LF. 
Therefore, our main results are not affected by whether these X-ray sources are \CIV$\lambda\lambda$1548,1550\,\AA\ contaminations. 
We perform the same Monte Carlo simulation as described in An14 to derive the detection completeness as a function of observed \Lya\ luminosity. 
The only difference is the assumption of EW distribution. 
The EW distribution of \Lya\ emission-line galaxies can be best fitted with an exponential law, d$N/d$EW = $N$ exp$^{-EW/\omega_{0}}$ \citep{Guaita10}, rather than the log-normal distribution for \Ha\ emitters. 
In this work, the best-fit exponential scale is $\omega_{0}$ = 38.3 $\pm$ 2.9 \AA\ with $\chi^{2}$ = 11.1, which is shown by the red line in Figure~\ref{f:ew.eps}.

\begin{figure}
\centering 
\includegraphics[width=0.45\textwidth]{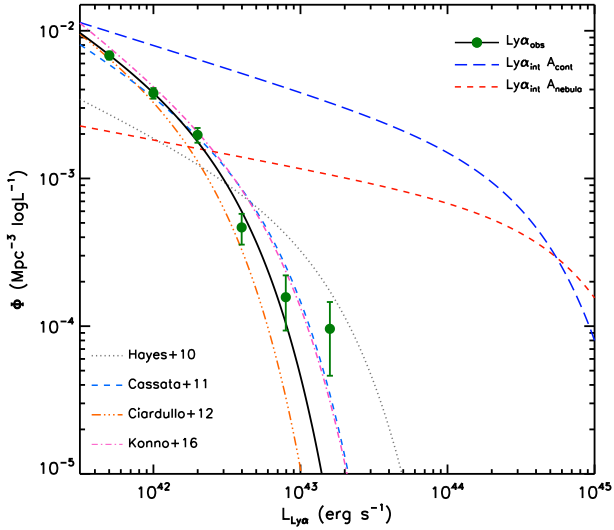}
\caption{Comparisons of \Lya\ and \Ha\ luminosity functions. The green solid dots are the \Lya\ LF data points based on our sample. 
The error bars are dominated by Poisson noise. Six AGNs are removed before we determine the LF. 
The black solid line shows the best-fit Schechter function based on our data points. The blue long-dashed line and red short-dashed line represent the intrinsic \Lya\ LFs derived from nebular extinction-($A\rm_{nebu}$) corrected \Ha\ LF and stellar continuum extinction-($A\rm_{cont}$) corrected \Ha\ LF, respectively. The \Lya\ LFs from previous works at a similar redshift are also plotted for comparison.} 
\label{f:global_esp.eps}
\end{figure}

The total coverage of 626\,arcmin$^{2}$ and the redshift span of $2.205<z<2.299$ correspond to a comoving volume of 1.29 $\times$ $10^{5}$\,Mpc$^{3}$. By dividing our sample into six bins in log($L_{\rm Ly\alpha}$) from 41.7 to 43.2 and taking the completeness correction, we obtained our \Lya\ LF data points as shown in Figure~\ref{f:global_esp.eps}. 
The error bars mainly represent the Poisson noise. We fit the data points with a Schechter function \citep{Schechter76} using the standard $\chi^{2}$ minimizer, providing the best-fit parameters $\Lstar$ = $10^{42.48 \pm 0.15}$\,erg\,s$^{-1}$, $\alpha$ = $-1.62\pm0.18$ , $\phistar = (1.16\pm0.64) \times 10^{-3}$\,Mpc$^{-3}$. We calculate the \Lya\ luminosity density by integrating our observed \Lya\ LF presented in Figure~\ref{f:global_esp.eps} (black solid line) and obtain $\rho_{obs}(L_{\rm Ly\alpha})$ = (8.12 $\pm$ 2.13) $\times 10^{39}$\,erg\,s$^{-1}$\,Mpc$^{-3}$. 

We update the \Ha\ LF from An14 with the improved extinction correction. The extra extinction correction for nebula, in effect, transfers the apparently faint galaxies to the brighter, shallower faint-end slope of \Ha\ LF with $\alpha$ = $-1.19\pm0.25$ (red short-dashed line in Figure~\ref{f:global_esp.eps}). The \Ha\ luminosity density does not change much because of the degeneracy of the three Schechter parameters. The updated intrinsic \Ha\ luminosity density is $\rho_{int}(L_{\rm H\alpha})$ = (2.52 $\pm$ 1.63) $\times 10^{40}$\,erg\,s$^{-1}$\,Mpc$^{-3}$. Then we determine the volumetric \Lya\ escape fraction using
\begin{equation} \label{e:v_e}
\begin{split}
fv_{esp} =\rho_{obs}(L_{\rm Ly\alpha})/(8.7 \times \rho_{int}(L_{\rm H\alpha})).
\end{split}
\end{equation}
We obtained the global \Lya\ escape fraction at $z$ = 2.24 to be (3.7 $\pm$ 1.4)\%.

We also calculate the intrinsic \Ha\ luminosity density by integrating the \Ha\ LF given in An14 to see the difference due to the update of extinction correction and obtain  
$\rho_{int}(L_{\rm H\alpha})$ = (3.30 $\pm$ 2.31) $\times 10^{40}$\,erg\,s$^{-1}$\,Mpc$^{-3}$. The consequent volumetric \Lya\ escape fraction based on $A\rm_{cont}$-corrected \Ha\ LF (blue long-dashed line in Figure~\ref{f:global_esp.eps}) is (2.8 $\pm$ 1.2)\%. Despite the large uncertainty due to the small sample size, we conclude that nebular extinction correction may cause a deviation of $\sim$ 1 percentage point in the global \Lya\ escape fraction. 

The \Lya\ LFs at $z = 2.2$ from previous studies are also shown in Figure~\ref{f:global_esp.eps} for comparison. Except for \cite{Hayes10a}, who has a shallower faint-end slope, our results agree well with the \Lya\ LFs from the literature. \cite{Konno16} derived a \Lya\ LF based on a sample of $> 3000$ LAEs from five independent blank fields, giving the best-fit Schechter parameters of $\Lstar$ = $10^{42.72 \pm 0.11}$\,erg\,s$^{-1}$, $\alpha$ = $-1.75^{+0.10}_{-0.09}$ , $\phistar = 6.32^{+3.08}_{-2.31} \times 10^{-4}$\,Mpc$^{-3}$. This more reliable LF is shown by the magenta dotted-dashed line in Figure~\ref{f:global_esp.eps}.

\section{DISCUSSION AND CONCLUSION} \label{s:discussion}
We have described a search for \Lya\ emitters at the same redshift as the \Ha\ emitters in the ECDFS from An14.
We identified 241 \Lya\ emitters with an excess in color $C393-NB393$. 
$C393$ is estimated from best-fit SEDs with MUSYC $UBVR$ imaging data. Six of the 241 are X-ray sources detected in the $Chandra$ 4\,Ms observations and all of them host an AGN in terms of X-ray luminosity. 
Although one of the six sources also has \Ha\ emission, we cannot exclude that they might be emitters of doublet lines of \CIV$\lambda\lambda$1548,1550\,\AA, which can also be detected through the NB393 filter. We fix the redshift of SEDs at $z=2.24$ to estimate the continuum in NB393. 
Although the \Lya\ absorption at this redshift will influence the estimate of the continuum, we reduced the influence by interpolating the continuum beside the absorption. We note that underestimate of the continuum has a negligible detrimental effect on our main results. From Figure~\ref{f:cdfs_mag_sele.eps}, we can see the scatter of $C393-NB393$ centered at zero is reasonably small. The EW cut is estimated from the dispersion envelope of $C393-NB393$ color. 
 
Among 124/241 LAEs in the overlapping area of 383 arcmin$^{2}$ in the ECDFS, only four sources are also \Ha\ emitters. One of the four LAHAEs is an X-ray source and classified as an AGN in \cite{Xue11}. Together with LAHAEs from COSMOS, we derive the AGN fraction in LAHAEs to be about 1/3.
\cite{Matthee16a} performed a detailed study of \Lya\ emission for \Ha\-selected galaxies at $z = 2.23$ in COSMOS and UDS fields and found 5/17 LAHAEs are AGNs.
Our fraction of AGNs is also similar to the AGN fraction among luminous \Ha\ emitters at $z \sim$ 0.8-2.23 \citep{Sobral16}. 
From Figure~\ref{f:luminosities.eps}, we can see that all of the AGNs in LAHAEs are also luminous \Ha\ emitters. 

We use the extinction-corrected \Ha\ luminosity to deduce the intrinsic \Lya\ luminosity and calculate the \Lya\ escape fraction. Naturally, the correction of dust attenuation for \Ha\ emitters is a key issue in our work. The Calzetti extinction law is used for the stellar continuum. Here, we updated the extinction correction for 56 \Ha\ emitters by employing the \cite{Cardelli89} Galactic extinction curve. Since the discrepancy between extinction coefficients at 6563\,\AA\ of these two curves is negligible ($< 0.01$\,mag), our main results on the \Ha\ emission are not affected. For the \Lya\ continuum at 1216\,\AA, the extinction coefficients ($k\rm_{Ly\alpha}$) of the Cardelli curve is 1.7 times that of Calzetti extinction law. It follows that the extinction-corrected \Lya\ luminosities for the four LAHAEs are largely affected. 
The other key issue in extinction correction is the relation between the color excess of the stellar continuum and nebular emission line. As confirmed by previous studies, the Calzetti's prescription with $E(B-V)\rm_{cont}$ = 0.44\,$E(B-V)\rm_{nebu}$ is no longer suitable for high-redshift SFGs \citep[e.g.,][]{Erb06,Wuyts13,Reddy15}. \cite{Wuyts13} suggested that an extra extinction correction is needed for high-$z$ \HII\ regions. We use their prescription to correct the extra extinction for both \Ha\ and \Lya\ emitters. As shown in Figure~\ref{f:extinction.eps}, the difference in the dust attenuation between ionized gas and stellar continuum increases with SFR. This strong correction was found by \cite{Reddy15} who did a detailed investigation of dust attenuation based on $z\sim 2$ SFGs using NIR spectroscopic data. 

The three LAHAEs in the COSMOS field are also corrected for SED-derived extinction with the Cardelli extinction law and an extra extinction for the nebular region as described above.
In total, five of the seven LAHAEs have extinction-corrected \Lya/\Ha\ ratio larger than the theoretical prediction of case B recombination. 
\cite{Song14} also found two of nine near-IR spectroscopically identified LAHAEs having intrinsic \Lya/\Ha\ ratio above that of case B recombination.
This could be due to a higher intrinsic \Lya/\Ha\ ratio in star-forming regions of high-redshift emitters than that of case B recombination. 
For example, the more complicated geometry and kinematics of the \HI\ region may enhance the emission of \Lya\ as explained by \cite{Song14}. 
It is worth noting that one of these four sources is an AGN. Another possibility is that the collisional excitation by shock from AGNs or supernova-derived winds contributed to the emission of \Lya. 
For high-redshift galaxies, extinction correction is one of the major sources of uncertainties. We stress that our estimate of dust attenuation is along the line of sight and limited by $A\rm_{V} \le 2.75$\,mag adopted by the dusty starburst template in EAZY \citep{Brammer08}. This apparently underestimates dust attenuation for galaxies with $A\rm_{V} > 2.75$.
It has been shown that the dust attenuation curve at high redshift is correlated with other properties, like stellar mass and galaxy spectral type \citep[e.g.,][]{Buat12, Zeimann15}. 
Although higher attenuation is suggested for nebular lines compared to the stellar continuum, we should note that the uncertainties remain large \citep[e.g.,][]{Forster09,Wuyts13, Reddy15}. Still, an anti-correlation is seen between the \Lya\ escape fraction and dust attenuation in Figure~\ref{f:ind_esp.eps}. In addition, an extra extinction correction for the \HII\ region leads to a steeper $f\rm_{esp}-E(B-V)$ correlation than the case where only continuum extinction  correction is adopted. 

To build the \Lya\ LF in the ECDFS at $z$ = 2.24, completeness as a function of \Lya\ luminosity is carefully estimated via Monte Carlo simulation and is applied to the observed \Lya\ LF. 
We removed the six X-ray sources to avoid the AGN contamination. 
The volumetric \Lya\ escape fraction derived from the extinction-corrected \Ha\ LF and observed \Lya\ LF in the ECDFS is (3.7 $\pm$ 1.4)\%. The way we used to determine the global \Lya\ escape fraction is to compare the observed \Lya\ and intrinsic \Ha\ luminosity densities. 
The advantage of this method is its independence from empirical correction for ionizing fluxes. 
However, this method is very sensitive to \Ha\ extinction correction.  Although the different extinction laws we used in this work and An14 do not change the \Ha\ LF and thus do not affect our main results, the variation in the nebular color excess of extinction leads to a discrepancy of $\sim$\,1 percentage point on the global \Lya\ escape fraction at $z=2.24$. Considering the small value of the \Lya\ escape fraction at this epoch \citep[$\le \sim 5\%$; e.g.,][]{Hayes10a,Matthee16a}, we cannot ignore this $\sim$\,1 percentage point deviation. There are large uncertainties in our results because the \Ha\ sample is small. Although larger surveys of \Ha\ and \Lya\ emitters at $z\sim 2$ have been carried out \citep[e.g.,][]{Lee12, Sobral13, Matthee16a}, from the observations and studies presented here, we make the case for appropriate extinction correction to be adopted for the ionized gas.

%appropriate extinction correction for ionized gas should be adopted.}

%In An14, we have demonstrated that the constant extinction correction for \Ha\ luminosity is not a good approximation. 
%The difference of the volumetric \Lya\ escape fractions between the two fields will decrease to $\sim$ 1 percentage point if we also adopt a constant extinction correction for our \Ha\ sample. 
%We also use an averaged \Lya\ LF derived from five different fields \citep{Konno16} to calculate the global \Lya\ escape fraction in ECDFS and COSMOS and the discrepancy also decrease to $\sim$ 1 percentage point. We stress that the discrepancy of the volumetric \Lya\ escape fractions is the combined effect of difference in extinction correction of \Ha\ LF and cosmic variance between the two fields. 
%Therefore, when comparing the global \Lya\ escape fractions at different cosmic times or studying their connections, such as the evolution of \Lya\ escape fraction \citep{Hayes11, Cassata15}, the discrepancy caused by cosmic variance should be taken into account, and the results derived from different fields at the same cosmic time should be averaged to reduce this effect.

\acknowledgments
We are grateful to the anonymous referee for a detailed report and valuable comments, which have improved the quality of this work. F.X.A. thanks Cong Ma for his kind assistance and thanks Ian Smail, A. Mark Swinbank, and Stijn Wuyts for the helpful discussion about extinction correction. 
This work is supported by the National Basic Research Program of China (973 Program 2013CB834900), National Natural Science Foundation of China through grant U1331110, and the Strategic Priority Research Program ``The Emergence of Cosmological Structures'' of the Chinese Academy of Sciences (grant No. XDB09000000). C.-N. H., J.-S. H. and X.-Y. X. acknowledge the support from the NSFC grant 11373027.

\end{document}